\documentclass[prbr,aps,twocolumn,groupedaddress,showpacs,letterpaper]{revtex4}

\usepackage{bm}
\usepackage{graphicx}
\usepackage{color}
\usepackage{ulem}
\usepackage[top=25truemm,bottom=8truemm,left=16truemm,right=19truemm]{geometry}

\begin{document}

\title{Observation of band crossings protected by nonsymmorphic symmetry in the layered ternary telluride Ta$_3$SiTe$_6$}
\author{Takafumi Sato,$^{1,2}$ Zhiwei Wang,$^3$ Kosuke Nakayama,$^1$ Seigo Souma,$^{2,4}$ Daichi Takane,$^1$ Yuki Nakata,$^1$ Hideaki Iwasawa,$^5$ Cephise Cacho,$^5$ Timur Kim,$^5$ Takashi Takahashi,$^{1,2,4}$ and Yoichi Ando$^3$}

\affiliation{$^1$Department of Physics, Tohoku University, Sendai 980-8578, Japan\\
$^2$Center for Spintronics Research Network, Tohoku University, Sendai 980-8577, Japan\\
$^3$Institute of Physics II, University of Cologne, K\"{o}ln 50937, Germany\\
$^4$WPI Research Center, Advanced Institute for Materials Research, Tohoku University, Sendai 980-8577, Japan\\
$^5$Diamond Light Source, Harwell Science and Innovation Campus, Didcot OX11 0DE, United Kingdom}

\date{\today}

\begin{abstract}
We have performed angle-resolved photoemission spectroscopy of layered ternary telluride Ta$_3$SiTe$_6$ which is predicted to host nodal lines associated with nonsymmorphic crystal symmetry. We found that the energy bands in the valence-band region show Dirac-like dispersions which present a band degeneracy at the $R$ point of the bulk orthorhombic Brillouin zone. This band degeneracy extends one-dimensionally along the whole $SR$ high-symmetry line, forming the nodal lines protected by the glide mirror symmetry of the crystal. We also observed a small band splitting near $E_{\rm F}$ which supports the existence of hourglass-type dispersions predicted by the calculation. The present results provide an excellent opportunity to investigate the interplay between exotic nodal fermions and nonsymmorphic crystal symmetry.
\end{abstract}


\maketitle

The search for new types of topological materials hosting nodal fermions is currently one of emergent topics in condensed-matter physics. Nodal fermions such as Dirac and Weyl fermions are characterized by the gapless low-energy excitation whose energy vs momentum ($k$) relation obeys the massless Dirac/Weyl equation. They provide a fertile ground to realize a variety of outstanding physical properties such as the extremely high mobility, large negative magnetoresistance, and chiral anomaly \cite{WanPRB2011, ZyuzinPRB2012, ChernodubPRB2014, AliNature2014, LiangNM2015, HuangPRX2015, XiongScience2016, HirschbergerNM2016, ZhangNC2017, LiNC2017}. In three-dimensional (3D) topological insulators, the nodal fermions manifest as a linearly dispersive spin-polarized Dirac-cone surface band whose degeneracy at the Dirac point is protected by time-reversal symmetry \cite{HasanReview, ZhangReview, AndoReview}.

One of the effective strategies to search for nodal fermions is to utilize the point-group symmetries of crystal, i.e. mirror reflection, rotation, and inversion symmetries in addition to time-reversal symmetry. This is highlighted by the discovery of topological crystalline insulators hosting the surface nodal fermions protected by mirror reflection symmetry \cite{FuPRL2011, HsiehNC2012, TanakaNP2012, XuNC2012, DziawaNM2012}. The mirror symmetry also plays an important role in the search for topological nodal-line semimetals (NLSMs) characterized by the continuous band crossing along a one-dimensional curve in the $k$ space (nodal line) \cite{Burkov2011, WuNP2016, BianNC2016, BianPRB2016, GuanSciAdv2016, WengPRB2016, YamakageJPSJ2016, TakaneQM2018}. When the rotational symmetry is taken into account, one can stabilize 3D Dirac semimetals such as Na$_3$Bi and Cd$_3$As$_2$ \cite{WangPRB2012, WangPRB2013, NeupaneNC2014, BorisenkoPRL2014, LiuScience2014}. Also, when the space inversion symmetry is broken, Dirac semimetals are transformed into Weyl semimetals with a spin-split pair of Weyl cones and Fermi-arc surface states \cite{XuScience2015, LvPRX2015, YangNP2015, SoumaPRB2016}.

While all these topological materials described above are based on the symmorphic symmetry which preserves the origin during the symmetry operation, the nonsymmorphic space-group symmetry combining the point-group symmetry and the fractional translation is recently attracting particular attention, because it further enriches the category of nodal fermions. This is demonstrated by the prediction/observation of nodal loops in NLSMs protected by the glide mirror (mirror reflection plus translational) symmetry \cite{XuPRB2015, SchoopNC2016, TakanePRB2016, NeupanePRB2016, LouPRB2016, ChenPRB2017}, as well as the Weyl nodes in trigonal Te and the nodal lines in ZrSiS, both of which are protected by the screw (rotation plus translational) symmetry \cite{SchoopNC2016, ChenPRB2017, NakayamaPRB2017}. Moreover, the nonsymmorphic symmetry is predicted to give rise to more exotic nodal-fermion features such as the M\"{o}bius-twist surface states \cite{ShiozakiPRB2015}, hourglass fermions \cite{WangNature2016, LiPRB2018}, and nodal chains \cite{BzduNature2016, WangNC2017, YanarXiv2017}. However, in contrast to these attractive predictions, the experimental proof of nodal fermions dictated by nonsymmorphic symmetries is still very limited \cite{MaSciAdv2017}. It is thus of great importance to experimentally establish new nodal-fermion materials protected by nonsymmorphic symmetries.

Recently, it was theoretically proposed that layered ternary telluride Ta$_3$SiTe$_6$ hosts the nodal fermions protected by the nonsymmorphic glide mirror symmetry \cite{LiPRB2018}. This material crystalizes in the orthorhombic structure with the space group No. 62 ($Pnma$). As shown in Fig. 1(a), the basic structural unit of Ta$_3$SiTe$_6$ is a Te trigonal prismatic slab with Ta atoms located around the center of this prism \cite{OhnoJSSC1999}. Each unit cell contains two such slabs which are overlaid with each other by the inversion operation. First-principles band-structure calculations \cite{LiPRB2018} show that, when the spin-orbit coupling (SOC) is neglected, Ta$_3$SiTe$_6$ displays a four-fold-degenerate (eight-fold-degenerate if counting spin) nodal line on the $SR$ line in the bulk Brillouin zone (BZ) [see Fig. 1(b)] due to the band crossing protected by the glide mirror symmetry. It is also suggested that when the SOC is included, the four-fold degeneracy on the $SR$ line is slightly lifted and as a result the hourglass-like dispersions appear in the close vicinity of $E_{\rm F}$. To examine such intriguing predictions, it is highly desirable to experimentally establish the electronic band structure of Ta$_3$SiTe$_6$.

In this Rapid Communication, we report angle-resolved photoemission spectroscopy (ARPES) of Ta$_3$SiTe$_6$ single crystal. By utilizing the energy-tunable photons from synchrotron radiation, we have determined the band structure in the 3D BZ, and found that the valence-band dispersion in the energy range much wider than the energy scale of SOC is characterized by the nodal lines on the SR line, whereas the band dispersion within 50 meV of $E_{\rm F}$ is substantially influenced by the finite SOC. We discuss implications of our observations in comparison with the first-principles band-structure calculations, and discuss the characteristics of the exotic nodal fermions in relation to the nonsymmorphic symmetries.

\begin{figure}
\begin{center}
 \includegraphics[width=3.4in]{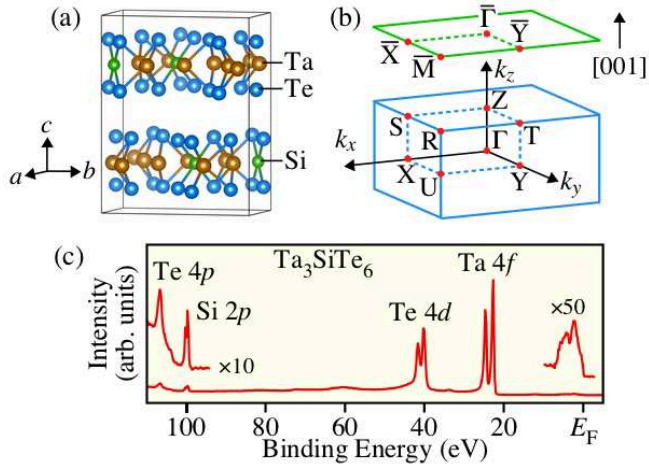}
\vspace{-0.5cm}
\caption{(color online). (a) Crystal structure of Ta$_3$SiTe$_6$. (b) Bulk orthorhombic BZ (blue) and corresponding surface BZ projected onto the (001) plane (green). (c) EDC of Ta$_3$SiTe$_6$ in a wide energy range measured at $h\nu$ = 600 eV. }
\end{center}
\end{figure}

High-quality single crystals of Ta$_3$SiTe$_6$ were grown by the chemical vapor transport method by using I$_2$ as transport agent. High-purity powders of Ta (99.99\%), Si (99.99\%) and Te shot (99.9999\%) were sealed in an evacuated quartz tube, which was subsequently put in a two-zone tube furnace. The temperatures in the furnace were set to be 950 $^\circ$C (source side) and 850 $^\circ$C (growth side), which were kept for 10 days. Note that the surface cleaning of Ta powders is crucial for obtaining large Ta$_3$SiTe$_6$ crystals; hence, we performed the surface cleaning procedure at 500 $^\circ$C using H$_2$ in a sealed quartz tube \cite{Wang1,Wang2}. ARPES measurements were performed with Omicron-Scienta R4000 and SES2002 electron analyzers with energy-tunable synchrotron-radiation light at the beamline I05 of Diamond Light Source and BL28 of Photon Factory (KEK). To excite photoelectrons, we used linearly polarized vacuum-ultraviolet (VUV) light of 40-100 eV. The energy and angular resolutions were set to be 6-20 meV and 0.2$^\circ$, respectively. We also measured the core-level spectrum with 600-eV photons at BL2 beamline of Photon Factory. Crystals were cleaved {\it {in-situ}} along the (001) plane in an ultrahigh vacuum of 1 $\times$ 10$^{-10}$ Torr and showed a shiny mirror-like surface indicative of high quality of the crystal. Sample temperature during measurements was kept at $T$ = 30 K. Figure 1(c) displays the energy distribution curve (EDC) in a wide energy region measured at photon energy ($h\nu$) of 600 eV. Several sharp core-level peaks are observed at the binding energy ($E_{\rm B}$) of 110, 100, 41, and 23 eV, which are attributed to the Te 4$p$, Si 2$p$, Te 5$d$, and Ta 4$f$ orbitals, respectively. Besides these core-level peaks, a weak structure originating from the valence band is resolved within 10 eV of $E_{\rm F}$.

\begin{figure}
 \includegraphics[width=3.4in]{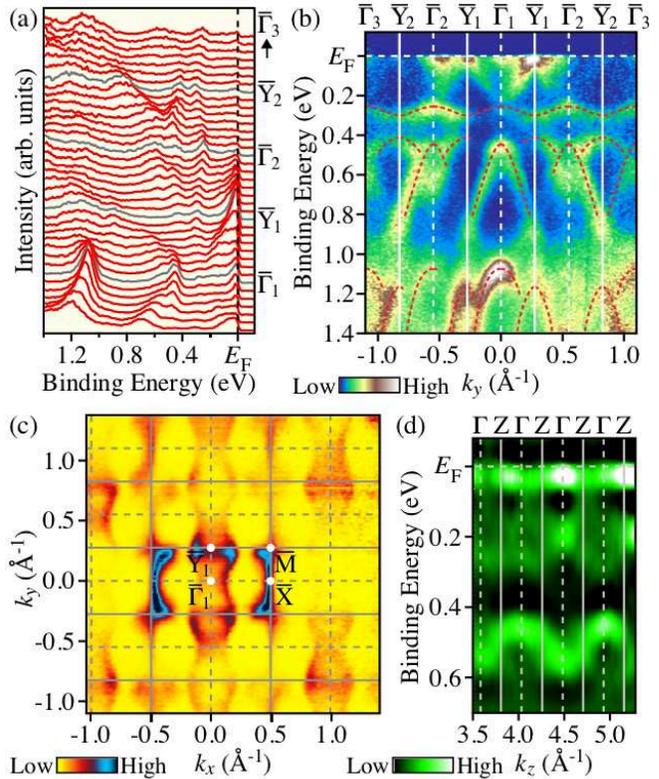}
\vspace{-0.5cm}
\caption{(color online).  (a) EDCs of Ta$_3$SiTe$_6$ at $T$ = 30 K measured along the $\bar \Gamma$$\bar Y$ cut with 87-eV photons. (b) Corresponding ARPES-intensity plot as a function of $E_{\rm B}$ and $k_y$. (c) ARPES intensity at $E_{\rm F}$ plotted as a function of in-plane wave vectors ($k_x$ and $k_y$) measured at $h\nu$ = 87 eV. Intensity at $E_{\rm F}$ was obtained by integrating the spectral intensity within $\pm$10 meV of $E_{\rm F}$. (d) Plot of normal-emission ARPES intensity as a function of $k_z$. Inner potential was estimated to be $V_0$ = 11.6 eV from the periodicity of the band dispersion.}
\end{figure}

\begin{figure*}
\includegraphics[width=6.6in]{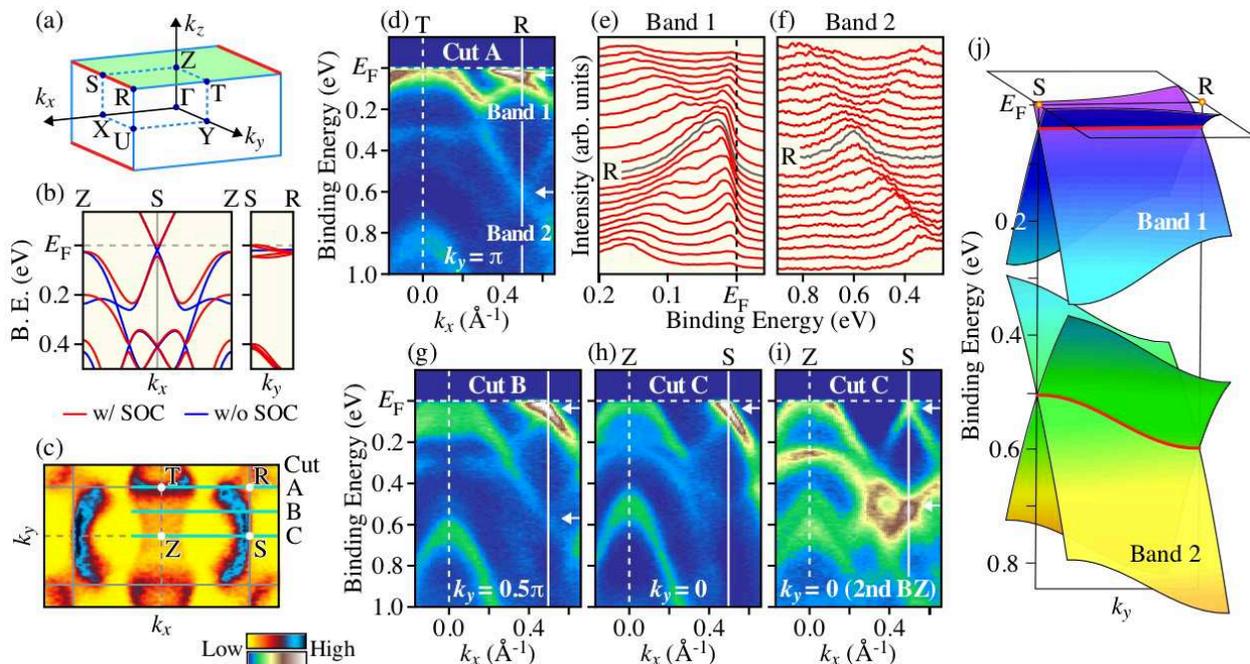}
\vspace{-0.5cm}
\caption{(color online). (a) Bulk BZ of Ta$_3$SiTe$_6$. The $k_z$ = $\pi$ plane ($SRTZ$ plane) and $SR$ line are indicated by green shade and red line, respectively. (b) Calculated band dispersions with (red curves) and without (blue curves) SOC \cite{LiPRB2018} along the $SZ$ (left) and $SR$ (right) cuts. (c) ARPES intensity at $E_{\rm F}$ plotted as a function of in-plane wave vectors measured at $h\nu$ = 87 eV. Intensity at $E_{\rm F}$ was obtained by integrating the spectral intensity within $\pm$10 meV of $E_{\rm F}$. (d) Plot of ARPES intensity measured along the $TR$ cut [cut A in (c)]. (e) and (f), EDCs around the $R$ point to highlight the band crossing for bands 1 and 2, respectively. (g)-(h) Same as (d) but measured along cuts B and C in (c), respectively. (i) Same as (h) but measured in the second BZ. Band-crossing points are indicated by white arrows in (d) and (g)-(i). (j) Schematic view of nodal lines on the $SR$ line based on the experimental valence-band dispersion.}
\end{figure*}

To see the overall valence-band structure, we show in Fig. 2(a) the EDCs at $h\nu$ = 87 eV measured over a wide $k$ region along the $\bar \Gamma$ $\bar Y$ cut of the surface BZ [see Fig. 1(b)]. One can identify several dispersive bands within 1.2 eV of $E_{\rm F}$, e.g., two holelike bands topped at 0.45 and 1.1 eV, respectively, at the $\bar \Gamma$ point of the first BZ ($\bar \Gamma$), and a sharp peak in the vicinity of $E_{\rm F}$ at around the $\bar \Gamma_1$ and $\bar \Gamma_2$ points. These energy bands are attributed to the Ta 5$d$ orbitals hybridized with the Te 4$p$ orbitals, with a negligible contribution from the Si 2$p$ states \cite{LiPRB2018}. To see more directly the band dispersions, we plot in Fig. 2(b) the ARPES intensity as a function of $k_y$ and $E_{\rm B}$, where we observe more clearly two holelike bands topped at 0.45 and 1.1 eV at $\bar \Gamma$ and a weakly dispersive band at $\sim$ 0.3 eV.  All of these bands well follow the periodicity of the surface BZ as indicated by red dashed curves, although the intensity is strongly modulated by the matrix-element effect of photoelectron intensity.  We also observe some dispersive bands within 0.2 eV of $E_{\rm F}$ between the $\bar \Gamma_1$ and $\bar \Gamma_2$ points, which cross $E_{\rm F}$ and form the Fermi surface.

To gain further insight into the Fermi-surface topology of Ta$_3$SiTe$_6$, we show in Fig. 2(c) the ARPES-intensity mapping at $E_{\rm F}$ as a function of in-plane wave vectors ($k_x$ and $k_y$). One can immediately recognize that the intensity pattern follows well the periodicity of the BZ over a wide $k_x$-$k_y$ area. We observe two types of Fermi surfaces, one is an open Fermi surface with a holelike character, surrounding the $\bar \Gamma$ and $\bar Y$ points with strong wiggling along the $k_x$ direction. Another Fermi surface encloses the $\bar M$ and $\bar X$ points, forming an open Fermi surface with an electronlike character. This Fermi surface is associated with the nodal line situated slightly below $E_{\rm F}$ as detailed later. The observed strong anisotropy of Fermi surfaces is consistent with the in-plane anisotropy of the crystal structure due to the large (about twice) difference in the lattice constant between the $a$- and $b$-axes.

To clarify the three-dimensionality of the electronic states, we investigated the band structure along the wave vector perpendicular to the surface ($k_z$) by varying the photon energy in the ARPES measurements. The ARPES-intensity plot as a function of $k_z$ in Fig. 2(d) signifies a few energy bands that clearly display a finite $k_z$ dispersion, e.g. two bands at $E_{\rm B}$ = 0.1-0.3 eV and 0.4-0.6 eV, respectively. This suggests that the observed energy bands are of bulk origin. Interestingly, these bands appear to exhibit a periodicity twice as large as that of the bulk BZ; for example, the band located at 0.4-0.6 eV has a top of dispersion at the $\Gamma$ point of $k_z$ = 4.0 {\text \AA} while it has a bottom at the adjacent $\Gamma$ point of $k_z$ = 4.5 {\text \AA} (note that it is difficult to trace the periodicity of the bands which cross $E_{\rm F}$ due to the small $k_z$ dispersion). This unexpected finding implies that the periodic potential from the unit cell consisting of the two monolayers of Ta$_3$SiTe$_6$ [see Fig. 1(a)] is rather weak, and electrons in the crystal actually feel more strongly the periodic potential from the one monolayer unit (see Supplemental Material for detailed comparison of the experimental band structure with the band calculations for monolayer \cite{SM}).

Since the $k_z$ value is experimentally established from the $h\nu$-dependent ARPES measurements, next we search for the predicted band crossing by selecting $k_z$. It has been suggested by the first-principles band-structure calculations \cite{LiPRB2018} that when the SOC is neglected, all the bands that cross the $SR$ line [red line in Fig. 3(a)] must be degenerate at the $k$ point that lies on the $SR$ line due to the crystal symmetry. This is clearly seen in the calculated band structure along the $ZSZ$ line shown by blue curves in Fig. 3(b), where the energy bands intersect with each other at the $S$ point and exhibit a characteristic X-shaped Dirac-like dispersion at the $S$ point. This degeneracy occurs on the entire $SR$ line, as seen from the right panel of Fig. 3(b), where the number of bands along the $SR$ line (two in this case) is unchanged along the $SR$ line, indicative of the robust degeneracy of bands. Upon inclusion of the SOC, the band degeneracy is lifted at the band-crossing points, as can be seen from a large band repulsion at  $E_{\rm B}$ = 0.25 eV at the midway between $Z$ and $S$ [note that our data in Fig. 3(i) are clearly in qualitative agreement with the calculation with SOC in this respect]. Also,  a finite spin-orbit gap opens at the $S$ point ($\sim$ 30 and $\sim$ 10 meV for the near-$E_{\rm F}$ band and the 0.4-eV bands, respectively), although its energy scale is much smaller than the band repulsion at $E_{\rm B}$ = 0.25 eV.

To have a deeper insight into the feature of the bands on a high-symmetry plane, we fixed the photon energy to 87 eV that can trace the $k_z$ $\sim$ $\pi$ plane ($SRTZ$ plane) [green shade in Fig. 3(a)]. Figure 3(d) displays the plot of ARPES intensity measured along cut A in Fig. 3(c) that corresponds to the $TR$ cut in the BZ. One can immediately identify two X-shaped bands with the intersection on the $R$ point at $E_{\rm B}$ $\sim$ 0.03 and 0.6 eV, respectively, as marked by white arrows (we label these two bands bands 1 and 2, respectively). To further clarify whether these bands have intersection, we surveyed the EDCs [Figs. 3(e) and 3(f)]. As for band 1 [Fig. 3(e)], two peaks (one near $E_{\rm F}$ and another at around 0.17 eV) are observed in the EDCs measured at the ${\bm k}$ points far away from the $R$ point [top and bottom EDCs in Fig. 3(e)] and they gradually merge into a single peak on approaching the $R$ point. A similar trend is also visible for band 2 [Fig. 3(f)]. These systematic variations of EDCs demonstrate the band crossing at the $R$ point. 

We further surveyed the band crossing along several cuts on the $SR$ line. Figures 3(g)-3(i) show the results of three representative cuts [cuts B and C in the first BZ as shown in Fig. 3(c) and cut C in the second BZ not shown in Fig. 3(c)], where we always observe X-shaped bands despite a clear change in the overall band structure upon variation of $k_y$. This suggests that, in this energy scale, the band-crossing point (intersection) can be regarded as continuously connected in the $SR$ direction, giving rise to two characteristic nodal lines as illustrated in Fig. 3(j). We found that the intersection (nodal point) of band 1 is consistently located at $E_{\rm B}$ = 20-30 meV irrespective of the measured ${\bm k}$ cuts (compare white arrows for band 1 among cuts A-C). On the other hand, the intersection of band 2 moves from $E_{\rm B}$ = 0.6 eV at the $R$ point [cut A, Fig. 3(d)] to 0.5 eV at the $S$ point [cut C, Fig. 3(i)], showing a dispersion of the nodal point [see the illustration based on the experimental band dispersion in Fig. 3(j)]. This trend in the dispersive feature is well reproduced in the calculations with SOC (and also without SOC) [right panel of Fig. 3(b)] where the energy bands at 0.4 eV at the $S$ point rapidly disperse downward to higher $E_{\rm B}$ on approaching the $R$ point while another bands near $E_{\rm F}$ are less dispersive. These results suggest that as long as the energy region of interest is much wider than the scale of SOC, one can characterize overall valence-band feature by the presence of two nodal lines. However, we will show below that the SOC actually lifts the band degeneracy of the nodal line near $E_{\rm F}$, by the measurements with an energy resolution much better than the scale of the calculated spin-orbit gap.

\begin{figure}
\includegraphics[width=3.4in]{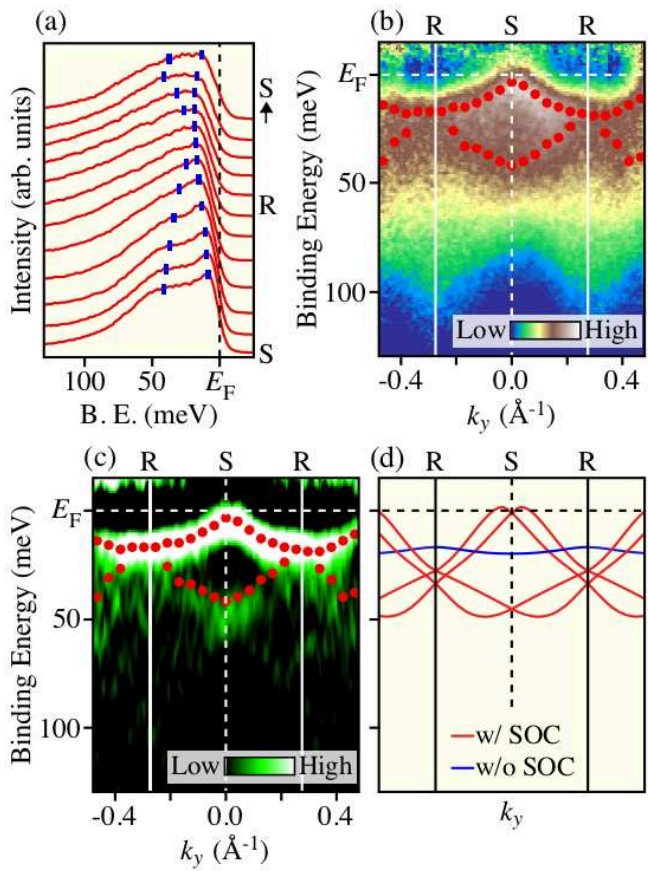}
\vspace{-0.5cm}
\caption{(color online). (a)-(c) EDCs, ARPES intensity, and second-derivative intensity in the vicinity of $E_{\rm F}$, respectively, measured along the $SR$ cut. Intensities in (b) and (c) are based on the EDCs divided by the FD function at 30 K convoluted with the resolution function. Red circles in (b) and (c) show the position of peaks determined by tracing the peak position of second-derivative of EDCs divided by the FD function. (d) Calculated band dispersions along the $SR$ cut with and without SOC (red and blue curves, respectively) \cite{LiPRB2018}.}
\end{figure}

Figure 4(a) shows the EDCs in the vicinity of $E_{\rm F}$ measured along the $SR$ cut with a higher energy resolution (6 meV). As visible from this figure, the overall spectral feature is not as sharp as one would expect from a single peak originating from the nodal line. This broad feature is not due to the energy resolution but may be ascribed to the $k_z$ broadening which is often the case for ARPES measurements with VUV photons. As shown in Fig. 4(a), the EDCs are composed of two peaks, e.g., at $\sim$ 10 and 40 meV, respectively, at the $S$ point. To estimate the intrinsic band energy, we divided the measured EDCs by the Fermi-Dirac distribution (FD) function at $T$ = 30 K convoluted with the resolution function, and show the results (the intensity and the second-derivative intensity) in Figs. 4(b) and 4(c). Now one can see more clearly two dispersive bands with the periodicity of the BZ. One band located closer to $E_{\rm F}$ shows a downward energy dispersion from the $S$ to $R$ point, while another band located at higher $E_{\rm B}$ exhibits an upward dispersion from the $S$ to $R$ point (red circles). This characteristic dispersive feature of the two experimental bands is not satisfactorily explained by the band calculations without the SOC [blue solid curve in Fig. 4(d)]. In contrast, the observed spectral feature looks compatible with the calculations including the SOC that predict hourglass-like dispersions along the $SR$ line in Ta$_3$SiTe$_6$ [red solid curves in Fig. 4(d)]. It is noted that the experimental band-crossing point near the $R$ point is closer to $E_{\rm F}$ than that in the calculation.

Now we discuss the characteristics of observed nodal lines. It was theoretically suggested that the nodal lines on the $SR$ line are solely dictated by the nonsymmorphic symmetry. When the SOC is absent, the $SR$ line is invariant under three key operations involving two glide mirrors [$\tilde{M}_x$$: (x, y, z)$ $\to$ $(-x+1/2, y+1/2, z+1/2)$ and $\tilde{M}_y$$: (x, y, z)$ $\to$ $(x+1/2, -y+1/2, z)$] and one mirror [$M_z$$: (x, y, z)$ $\to$ $(x, y, -z+1/2)$], leading to the four degenerate orthogonal states (including original Bloch state) and the resultant four-fold degenerate nodal lines (eight-fold when spin is counted) \cite{LiPRB2018}. Since this symmetry protection is valid for all energy bands, one can realize multiple (two) nodal lines stemming from bands 1 and 2 on the $SR$ line in the experiment as shown in Fig. 3(j). It is emphasized that the four-fold-degenerate nodal line is distinct from the nodal lines realized in other NLSMs such as ZrSiS, CaAgAs, and PbTaSe$_2$ which show the two-fold degeneracy \cite{BianNC2016, BianPRB2016, YamakageJPSJ2016, TakaneQM2018, SchoopNC2016, TakanePRB2016, NeupanePRB2016, LouPRB2016, ChenPRB2017}. As shown in Fig. 4(d), the nodal line is expected to split into four energy bands after inclusion of the SOC, forming an hourglass dispersion centered at the $R$ point. This theoretical prediction is in good agreement with the present high-resolution ARPES results [Figs. 4(a)-4(c)]. Since the observed energy bands forming nodal line/hourglass dispersions are located in a narrow energy range near $E_{\rm F}$ and would largely contribute to the density of states, the nodal-fermion-related exotic physical properties such as nontrivial Berry phase \cite{AnPRB2018} may be realized by doping a small amount of holes (i.e. by moving the nodal line or the necks of the hourglass dispersions to $E_{\rm F}$ so that they affect more strongly the transport properties) $via$, e.g., the chemical substitution in the crystal. We point out though that such realization is strongly influenced by another Fermi surface located at the $\bar \Gamma$ point [Fig. 2(c)].

In conclusion, we reported high-resolution ARPES results on layered ternary telluride Ta$_3$SiTe$_6$. We have revealed the existence of two nodal lines near $E_{\rm F}$ and at $\sim$ 0.5 eV, respectively, on the $SR$ line in the 3D BZ, consistent with the first-principles band-structure calculations without the SOC. The observed two nodal lines are protected by the nonsymmorphic glide mirror symmetry of the crystal. We also observed a small band splitting near $E_{\rm F}$ which supports the existence of hourglass-type dispersions predicated by the calculation. The present result provides an excellent platform to study the role of nonsymmorphic symmetry to the formation of nodal fermions.

\begin{acknowledgements}
We thank K. Nakamura, H. Oinuma, K. Shigekawa, K. Hori, K. Horiba, and H. Kumigashira for their assistance in the ARPES measurements. We also thank Diamond Light Source for access to beamline I05 (proposal number S18839). This work was supported by MEXT of Japan (Innovative Area ``Topological Materials Science'', JP15H05853), JSPS (KAKENHI No: JP17H01139, JP26287071, JP25220708, JP16K13664, JP16K17725, and JP18H01160), KEK-PF (Proposal number: 2018S2-001, 2016G-555), and DFG (CRC1238 ``Control and Dynamics of Quantum Materials'', Project A04).
\end{acknowledgements}

\bibliographystyle{prsty}

\end{document}